\documentstyle[aps,prb,twocolumn,psfig]{revtex}

\begin{document}
\title{Hole spin polarization in GaAlAs:Mn structures}
\author{A. Ghazali}
\address{Groupe de Physique des Solides,\\
UMR 7588-CNRS, Universit\'{e}s Paris 7\\
et Paris 6\\
Tour 23, 2 Place Jussieu, F-75 251 Paris Cedex 05, France}
\author{I. C. da Cunha Lima, and M. A. Boselli}
\address{Instituto de F\'\i sica, Universidade do Estado do Rio de Janeiro\\
Rua S\~{a}o Francisco Xavier 524, 20.500-013 Rio de Janeiro, R.J., Brazil}
\maketitle
\date{\today}

\begin{abstract}
A self-consistent calculation of the electronic properties of GaAlAs:Mn
magnetic semiconductor quantum well structures is performed including the
Hartree term and the sp-d exchange interaction with the Mn magnetic moments.
The spin polarization density is obtained for several structure
configurations. Available experimental results are compared with theory.
\end{abstract}


During the last years spin-polarized transport became a matter of major
interest \cite{nat1,nat2}, giving rise to a field usually called
magnetoelectronics, or spintronics. Much effort has been made in studying
and producing the so-called spin-valve, where a spin-polarized current is
generated. A promising system in that area is the recently grown structure
of GaAs/AlAs with inclusions of Ga$_{1-x}$Mn$_{x}$As layers \cite
{van1,van2,oiwa1,oiwa2,matsu,ohno1,ohno2}. Homogeneous samples of
Ga$_{1-x}$Mn $_{x}$As alloys with $x$ up to $7\%$ have been produced,
avoiding the formation of MnAs clusters by using low temperature
($200-300^{o}$ C) MBE techniques. In Ga$_{1-x}$Mn$_{x}$As, a
new prototype of Diluted Magnetic Semiconductors (DMS) \cite{dms1,dms2},
the Mn$^{2+}$ cations have the $3d$
shell partially filled with five electrons, in such a way that they carry a
magnetic moment of $S=5/2$. In addition, the Mn ion binds a hole to satisfy
charge neutrality. Besides its practical importance, this kind of DMS
introduces an interesting problem from the physical point of view: Mn in the
alloy is a strong $p$ dopant, the free hole concentration reaching even
$p=10^{20-21}cm^{-3}$. At small Mn concentrations, the alloy is a
paramagnetic insulator. As $x$ increases it becomes ferromagnetic, going
through a non-metal/metal transition for higher concentrations ($x\approx
0.03$), and keeping its ferromagnetic phase. For $x$ above $5\%$, the alloy
becomes a ferromagnetic insulator. In the metallic phase, the ferromagnetic
transition is observed in the range of $30-100$K, depending on the value of
$x$, among the highest transitions temperatures observed in DMS. The
{\it sp-d} exchange interaction of Ruderman-Kittel-Kasuya-Yosida (RKKY)
type has been recognized as the main origin of the observed ferromagnetism
in the metallic phase of III-V based DMSs \cite{matsu}.

A few years ago Helman and Baltensperger \cite{helman1,helman2} performed
analytical RKKY calculations for the spin polarization in several confined
structures, and clarified the role of the confined states in the multilayer
interaction. Recently, improvements on the standard RKKY mechanism were
introduced by Byounghak Lee {\it et al} \cite{lee} to treat the ferromagnetism
of GaMnAs quantum wells.
In the present work we consider a Ga$_{1-x}$Mn$_{x}$As layer in
its metallic phase, grown inside a GaAs/AlAs quantum well structure. We
obtain the spin-polarized electronic structure for holes, taking into
account their interaction with the magnetic impurities. The Hamiltonian we
consider is:

\begin{equation}
H=H_{0}+U_{H}+U_{mag},  \label{hamil}
\end{equation}

\begin{equation}
U_{mag}(\vec{r})=-I\sum_{i=1}^{N_{i}}\vec{s}(\vec r).\vec{S}(\vec{R_{i}})
\delta (\vec{r}-\vec{R_{i}}).  \label{umag}
\end{equation}

The $H_{0}$ term in the rhs of Eq. (\ref{hamil}) contains the kinetic energy
and the confining potential, $U_{c}(z)$, due to the band edges mismatch at
the semiconductor interfaces. Coulomb interactions between carriers are
considered through the Hartree term $U_{H}\left( \vec{r}\right) $. The hole
system is supposed to be homogeneous in the $xy$ plane, so
$U_{H}(\vec{r})=U_{H}(z).$ In the present approach we treat
the magnetic interaction as
being due to an uniform magnetization in the DMS. If a net magnetization
exists, it will polarize the hole gas. This problem is solved
self-consistently by a secular matrix equation in the reciprocal space. The
method would be exact were not for cutting the matrix size. The advantage is
that it provides spin-polarized eigenvalues and eigenfunctions with high
accuracy, not only for bound states, but also for a high number of
scattering states. For each spin, we define the wavefunction Fourier
Transform (FT):

\begin{equation}
\psi _{\sigma }(\vec{r})=\int d^{3}q\exp (i\vec{q}.\vec{r})
\psi _{\sigma } (\vec{q}).  \label{psiq}
\end{equation}
The hole eigenstates will be obtained by discretizing the integrals on
$\vec{q}$ appearing in 
\begin{equation}
\int d^{3}r\psi _{\sigma }^{\ast }(\vec{r})(H-E)\psi _{\sigma }(\vec{r})=0.
\label{expect}
\end{equation}
When integrating the magnetic term in the Hamiltonian over $\vec{r}$, we
assumed the magnetic impurities to be uniformly distributed in the
Ga$_{1-x}$Mn$_{x}$As DMS layer, all of them having the same magnetization,
namely the thermal average magnetization $<\vec{M}>$. This treatment
includes not only the ferromagnetic phase but also phases where a partial
magnetization is observed, the ``canted-spin'' phases \cite{bosel1,bosel2}.
Therefore,
\begin{eqnarray}
-I\int d^{3}r\exp [i(\vec{q}-\vec{q}^{\prime }).\vec{r}]\sum_{i=1}^{N_{i}} 
\vec{s}(\vec{r}).\vec{S}(\vec{R_{i}})\delta (\vec{r}-\vec{R_{i}}) &\simeq&
\nonumber \\ 
-I \frac{\sigma }{2}<M>\sum_{i=1}^{N_{i}}\exp [i(\vec{q}-\vec{q}^{\prime
}). \vec{R_{i}}]=  \nonumber \\ 
-I\frac{\sigma }{2}<M>\frac{N_{i}}{V}\int d^{3}R_{i}\exp [i(\vec{q}-\vec{q}
^{\prime }).\vec{R_{i}}] &=&  \nonumber \\
-\widetilde{I}\frac{\sigma }{2}
x<M>F_{DMS}(q_{z}-q_{z}^{\prime })(2\pi )^{3} \delta ^{2}
(\vec{q}_{\parallel}-\vec{q}_{\parallel }^{\prime }),  \label{imag}
\end{eqnarray}
where $\frac{N_i}{V}$ is the impurity density, and $\sigma =\pm 1$ for spin
parallel (upper sign) or anti-parallel (lower sign) to the magnetization.
$F_{DMS}$ is the integral performed on the $z$ -coordinate along the DMS
layer: 
\begin{equation}
F_{DMS}(q)\equiv \frac{1}{2\pi }\int_{DMS}dz\exp [iq.z].  \label{form}
\end{equation}
In Eq.(\ref{imag}) we have defined $\widetilde{I}=\frac{I}{v_{0}}$, where
$v_{0}$ is the volume of the Mn$^{2+}$ ion, which is the $fcc$ lattice's
primitive cell volume, $a^{3}/4$. It is worthwhile to mention that, in
experimental works, $\widetilde{I}$ is usually represented by $N_{0}\beta$.
Here we used $N_{0}\beta=-1.2$eV \cite{okaba}. After performing the FT of
the Hartree term plus the confining potential, 
\begin{equation}
U_{el}(q)=\frac{1}{2\pi }\int dz\exp [iq.z][U_{H}(z)+U_{c}(z)],  \label{ftu}
\end{equation}
the \ eigenvalues and eigenfunctions at the bottom of the 2-D subbands may
be obtained by solving the secular matrix for each spin-polarization: 
\begin{eqnarray}
\det \left\{ \left[ \frac{\hbar ^{2}q_{z}^{2}}{2m^{\ast }}-E\right] \delta
(q_{z}-q_{z}^{\prime })+U_{el}(q_{z}-q_{z}^{\prime })- \right. \nonumber \\
\left. \widetilde{I}
\frac{\sigma }{2}<M>F_{DMS}(q_{z}-q_{z}^{\prime })\right\} =0.  \label{secul}
\end{eqnarray}

For applications, favorable spin configurations may be designed by growing a
proper structure. In particular, the possibility of having a ferromagnetic
order in the DMS heterostructures is important for spin tunneling and
resonant spin tunneling in nanostructures \cite{ohno1,bruno}. We have
calculated the electronic structure, and the spin polarization, in three
GaAlAs/GaAs:Mn quantum wells. We started with an AlAs/GaAs QW in the middle
of which a DMS barrier (Ga$_{0.65-x}$Al$_{0.35}$Mn$_{x}$As) is grown in its
ferromagnetic metallic phase. In that case, we have, actually, two QW of
widths $L$ separated by an internal barrier of 168.3 meV with thickness $d$,
and barriers of 529 meV at the lateral boundaries. In a second structure, we
assume that the DMS is just a Ga$_{1-x}$Mn$_{x}$As layer of thickness $d$,
no internal barrier being structurally imposed. Finally, in a last
structure, we explore the case of a single DMS QW: a Ga$_{1-x}$Mn$_{x}$As
layer of thickness $2L+d$ is surrounded by thick layers of AlAs. It is well
known that in ferromagnetic metallic Ga$_{1-x}$Mn$_{x}$As layers (in our
calculation we assumed $x=0.05$) the density of free carriers (holes) is
just a fraction of the density of the magnetic ions. Throughout the present
calculation we made the hole density $p=1.\times 10^{20}cm^{-3}$, T=0$K$,
and $<M>=5/2$. Due to the high carrier density, the Hartree term cannot be
neglected, and several subbands happen to be occupied.

The results for the energy of the bound states are shown in
Table \ref{tab01}, the arrows indicating spin up or down, i.e.,
parallel or anti-parallel to
the DMS magnetization. For the first two structures $L=50$ nm, and $d=1$ nm.
The criteria for self-consistency has been chosen to be a ratio $%
(E_{F}(i)-E_{F}(i-1))/2(E_{F}(i)+E_{F}(i-1))<.00001$, where $E_{F}(i)$ is
the calculated Fermi energy exiting iteration $i$.

\begin{figure}[tbp]
\psfig{figure=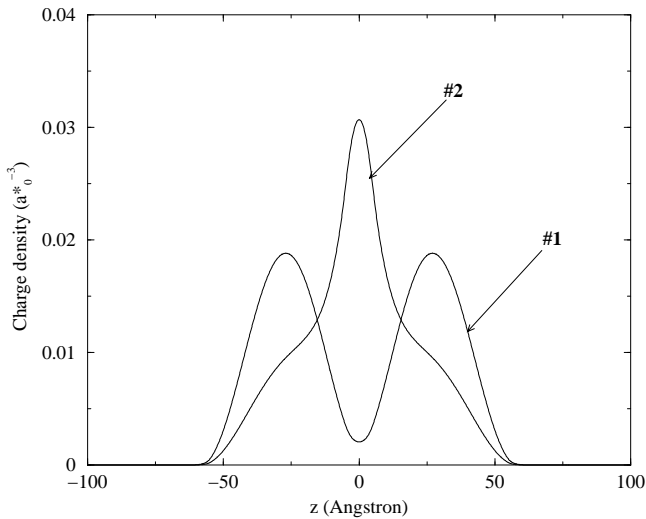,width=8.5cm}
\caption{Charge density as a function of the distance to the center of the
well. Sample \#1: two GaAs quantum wells of 50\AA \ separated by a 10\AA \
barrier of Ga$_{0.60}$Al$_{0.35}$Mn$_{0.05}$. Sample \#2:
a 10\AA \ Ga$_{0.95}$Mn$_{0.05}$As instead of the middle barrier.}
\label{fig01}
\end{figure}

\begin{figure}[tbp]
\psfig{figure=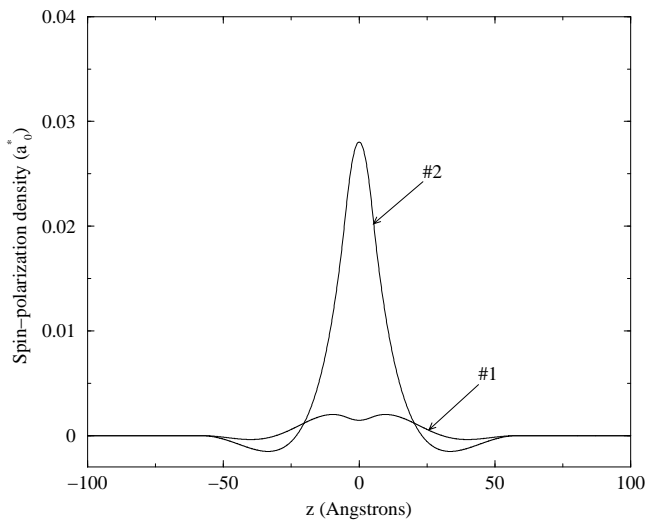,width=8.5cm}
\caption{Spin polarization density as a function of the distance to the
center of the well, for samples \#1 and \#2.}
\label{fig02}
\end{figure}

The charge density distributions are shown in Fig. \ref{fig01} for samples
\#1 and \#2. In the case of the first sample we observe a dip at the middle
of the structure, as a consequence of the existence of the central barrier.
A much less pronounced dip is observed in the sample \#1's spin polarization
density , shown in Fig. \ref{fig02}. By integrating the spin density we
obtain that 6\% of the spins are polarized.

\begin{table}[tbp]
\caption{Energies (in eV) of the bound states, reckoned from the top of the
AlAs valence band edge. Arrows represent the spin polarization of the state,
up for parallel, down for anti-parallel to the average Mn$^{2+}$
magnetization.}
\label{tab01}
\begin{tabular}{lcccccc}
state & sample 1 & spin & sample 2 & spin & sample 3 & spin \\ \hline
1 & -0.51819 & $\downarrow$ & -0.54815 & $\downarrow$ & -0.59511 & $\downarrow$ \\ 
2 & -0.51398 & $\uparrow$ & -0.51799 & $\uparrow$ & -0.58266 & $\downarrow$
\\ 
3 & -0.51244 & $\downarrow$ & -0.51233 & $\downarrow$ & -0.56052 & 
$\downarrow$ \\ 
4 & -0.51199 & $\uparrow$ & -0.51167 & $\uparrow$ & -0.52939 & $\downarrow$
\\ 
5 & -0.47744 & $\downarrow$ & -0.49945 & $\downarrow$ & -0.48941 & 
$\downarrow$ \\ 
6 & -0.46498 & $\uparrow$ & -0.47861 & $\uparrow$ & -0.44520 & $\uparrow$ \\ 
7 & -0.45885 & $\downarrow$ & -0.46045 & $\downarrow$ & -0.44070 & 
$\downarrow$ \\ 
8 & -0.45722 & $\uparrow$ & -0.45806 & $\uparrow$ & -0.43300 & $\uparrow$ \\ 
9 & -0.40581 & $\downarrow$ & -0.42870 & $\downarrow$ & -0.41125 & $\uparrow$
\\ 
10 & -0.38491 & $\uparrow$ & -0.40766 & $\uparrow$ & -0.38341 & $\downarrow$
\\ 
11 & -0.36957 & $\downarrow$ & -0.37390 & $\downarrow$ & -0.38069 & $\uparrow
$ \\ 
12 & -0.36612 & $\uparrow$ & -0.36903 & $\uparrow$ & -0.34152 & $\uparrow$
\\ 
13 & -0.30218 & $\downarrow$ & -0.32280 & $\downarrow$ & -0.31775 & 
$\downarrow$ \\ 
14 & -0.27688 & $\uparrow$ & -0.30397 & $\uparrow$ & -0.29390 & $\uparrow$
\\ 
15 & -0.24647 & $\downarrow$ & -0.25385 & $\downarrow$ & -0.24409 & 
$\downarrow$ \\ 
16 & -0.24076 & $\uparrow$ & -0.24624 & $\uparrow$ & -0.23809 & $\uparrow$
\\ 
17 & -0.16722 & $\downarrow$ & -0.18438 & $\downarrow$ & -0.17455 & $\uparrow
$ \\ 
18 & -0.14316 & $\uparrow$ & -0.16859 & $\uparrow$ & -0.16311 & $\downarrow$
\\ 
19 & -0.09378 & $\downarrow$ & -0.10369 & $\downarrow$ & -0.10420 & $\uparrow
$ \\ 
20 & -0.08594 & $\uparrow$ & -0.09386 & $\uparrow$ & -0.07647 & $\downarrow$
\\ 
21 & -0.01095 & $\downarrow$ & -0.02272 & $\downarrow$ & -0.03001 & $\uparrow
$ \\ 
22 & -- & -- & -0.01170 & $\uparrow$ & -- & -- \\ \hline
E$_F$ & -0.49486 & -- & -0.50324 & -- & -0.40430 & --
\end{tabular}
\end{table}

The elimination of the central barrier in sample \#2 increases the
probability of finding the holes in the middle of the structure, now
attracted by the unbalanced negative charges. The dip disappears giving rise
to a pronounced peak in both the charge density and the spin polarization
density. Another important difference is the Friedel-like oscillation in the
polarization, being anti-parallel in the middle of the well, and parallel
near the interfaces. The total polarization reaches 40\%.

\begin{figure}[tbp]
\psfig{figure=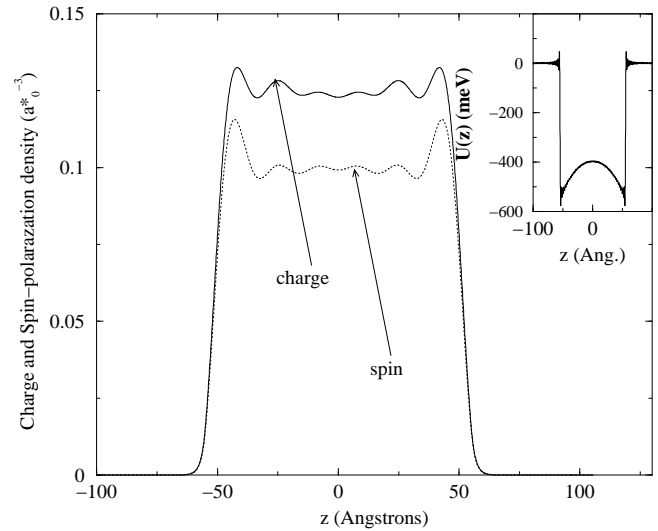,width=8.5cm}
\caption{Sample \#3: Charge density and spin density distributions in a
single quantum well of 110\AA \ Ga$_{0.95}$Mn$_{0.05}$As.
Inset: bare confining potential plus self-consistent Hartree term.}
\label{fig03}
\end{figure}

A radical change occurs when the DMS occupies the entire well, as shown in
Fig. \ref{fig03}. The number of carriers per unit area is much greater than
in the previous cases, and many subbands are occupied. The charge density
shows an oscillation inside the well, as a consequence of the higher
subbands occupation. The oscillations that occurs in the spin polarization
density follow those in the charge density. In the case where charges and
magnetic moments cohabit inside a finite width determined by the quantum
confinement, we obtain that the oscillations are not enough to invert the
polarization in any region. We plotted the sum of the bare confining
potential plus the self-consistent Hartree term in the inset of Fig. \ref
{fig03}. An effective central barrier appears at the middle of the well,
together with two valleys at the interfaces what increases the lateral
barriers from 529 meV to about 600 meV. In sample \#3 the magnetic
interaction between holes and the ferromagnetic layer contributes to the
confining potential with an increase of 150 meV for spin anti-parallel, and
a decrease of the same amount for spin parallel. Therefore, there is no
rigid shift of the eigenenergies as in a 3-D system. The total polarization
is 82\%.

The spin-polarized electronic structure in diluted magnetic semiconductors,
as obtained here, points to the possibility of a self-consistent calculation
of the magnetization in multilayered structures,
as in Refs. \onlinecite{bosel1} and
\onlinecite{bosel2}, taking into account a spin-polarized RKKY mechanism.
Recently Chiba {\it et al} \cite{chiba} investigated a GaMnAs trilayer
structure and observed a ferromagnetic, although weak, interaction
between two ferromagnetic layers.
The present result is
also important for determining properties like spin diffusion and spin
filtering in cases where spin-coherence lifetimes are large. For instance,
the spin motion for carriers polarized parallel and anti-parallel to the
equilibrium polarization in otherwise non-magnetic GaAs samples has been
predicted \cite{flatte} to show a difference of an order of magnitude
between the two speeds, with electrons and holes behaving in opposite way.
This fact points to the importance of taking a properly calculated
electronic structure in obtaining the spin-polarized transport in magnetic
GaMnAs structures. On the other hand, a spin filtering mechanism has been
previewed in the ZnSe/ZnMnSe/ZnSe structure with an external applied
magnetic field \cite{egues}. The present results on the electronic structure
of GaMnAs shows the possibility of filtering spins without needing an
external field in the range of temperatures of ferromagnetic phases.

\acknowledgements
This work was partially supported by CAPES, FAPERJ and CENAPAD/UNICAMP-FINEP
in Brazil, and by the PAST grant from Minist\`{e}re de l'\'{E}ducation
Nationale, de l'Enseignement Sup\'{e}rieur et de la Recherche (France).

\end{document}